# OntoFM: A Personal Ontology-based File Manager for the Desktop

Jenny Rompa, Giorgos Lepouras, Costas Vassilakis, and Christos Tryfonopoulos

Department of Computer Science & Technology,
University of Peloponnese, GR22100, Tripoli, Greece
{cst06035,gl,costas,trifon}@uop.gr

**Abstract.** Personal ontologies have been proposed as a means to support the semantic management of user information. Assuming that a personal ontology system is in use, new tools have to be developed at user interface level to exploit the enhanced capabilities offered by the system. In this work, we present an ontology-based file manager that allows semantic searching on the user's personal information space. The file manager exploits the ontology relations to present files associated with specific concepts, proposes new related concepts to users, and helps them explore the information space and locate the required file.

## 1 Introduction

As the number of files that a user stores in her computer grows, it is becoming increasingly difficult to manage stored documents. Personal Information Management (PIM) [2,9,10] aims at supporting users in the collection, storage and retrieval of personal information. In PIM systems, personal ontologies [8] can offer the semantic basis for managing the wealth of files saved in a user's local repository by relating user domain concepts to user folders and files. However, personal ontologies alone are not sufficient to manage the stored information, as users need specialized tools that leverage the personal ontology and help them to perform everyday tasks easily and effortlessly [4]. Other PIM approaches have also aimed at employing semantic web techniques to enhance the searching functionalities of the user's desktop [3]. In this demonstration, we present a novel file manager that bases its interactivity on the user's personal ontology, offering semantic browsing and searching mechanisms for locating files, a mind map inspired ontology visualization, and simple-to-use intuitive functionality that encourages less experienced users.

## 2 File Manager Prototype

### 2.1 Architecture

To implement the file manager prototype we opted for a modular architecture. We based our implementation on Protégé (http://protege.stanford.edu), an extensible open source ontology editor and knowledge-base framework. The file manager is implemented as a Protégé tab widget, retrieving information from a loaded personal ontology. Alternatives for the management, maintenance and population of the personal ontology may include automatic (e.g., metadata-based) and semi-automatic (like inheritance of a user-specified relationship for all files in a directory) approaches. Population and maintenance of the ontology is out of the scope of this paper; for the presented prototype we utilise an already populated one from [8]. Ontology visualization is based upon the OntoGraf (http://protegewiki.stanford.edu/wiki/OntoGraf) plug-in which has been extended to comply with the mind map paradigm [1]. Currently the ontology visualization pane permits only navigational functions, while other ontology management functions have been hidden to ease the complexity of the user interface. The notion of using ontology visualization methods as browsing aids has also been explored in [5,6].

## 2.2 Functionality

One of the most common functions of a file manager is to help users locate files in their personal file system. Depending on the information already known by the user we can categorize the file location task as (i) *perfect exploration*, where the user wants to locate a file when she knows the file name and location, and (ii) *limited/incomplete/partial exploration*, where the user wants to locate a file when she has incomplete information. Possible cases and file retrieval strategies for this file location task are:

- *Location is known*. If the file location is known the user can navigate to location and browse through files to find the one required. A file preview may help file browsing.
- *Filename is known*. If all or part of the filename is known the user can employ a search facility to scan the file system.
- *Metadata is known* (date, keywords from content, author, etc.). Current file managers provide mechanisms for searching for files with specific metadata as well as with specific content (full text search). The user can combine keyword search and criteria on other metadata such as file type or creation date. In some file managers, search criteria may be progressively refined to limit the number of results. However, search is not always easy. Search may present too many results, metadata may not filter results adequately, or the user may not even remember the exact keywords. To this end, ontology-based search may help alleviate search problems and facilitate file location.

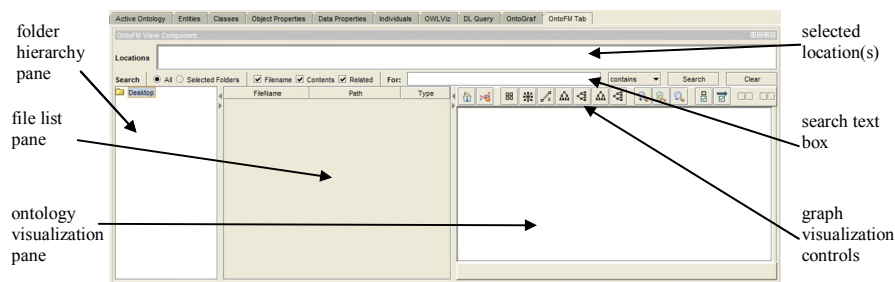

**Figure 1:** OntoFM initial screen with an explanation of the different components.

The user interface of the OntoFM file manager is presented in Figure 1. In the case that the user remembers the file location, she may use the folder hierarchy pane to browse through the folders and locate the file. Clicking on a folder name expands the folder hierarchy under the selected folder and reveals all subfolders of the first level as in a normal file manager. In the file list pane, only files of the folder selected are displayed and not files of subfolders. The user may exploit multiple-select functionality to select more than one folder and view all included files.

If the user does not remember the file location then she may use the search text box to enter keywords. As soon as the user starts typing in the text box, the file manager tool searches the ontology for instances that match what the user types and proposes candidate terms from the ontology. Using this type of search allows the user to recover files associated with a concept, although she does not remember their name or location, and extends file search and organization beyond naming conventions. Note also that since File is a concept in the ontology the search includes also filenames along with other ontology concepts. Once an ontology instance is selected, all concepts related with this file appear at the ontology visualization pane and may be used to further refine the search, as illustrated in Figure 2.

The ontology visualization pane shows terms entered, as well as instances directly related to each term. The user may expand a node by double clicking on it; upon expansion the related concepts of a node appear in the ontology visualization pane. Similarly the user may collapse a node by double clicking on it again. Additionally, the user may control the appearance of the ontology graph to center the graph on a concept, or to create a tree under or above the selected concept. The file list pane displays all file instances found in the ontology and are directly related to the entered key terms. Files displayed have at least one direct relation to a concept entered by the user. Not all concepts have to be related with all files displayed. However, files related to most concepts receive a higher score and appear on top of the list, providing a ranking of the retrieved files. The ontology visualization pane may be used for result refinement: the user may select a concept and enter a constraint for it, limiting the files shown in the file list pane to those being related to an instance of this concept that satisfies the constraint.

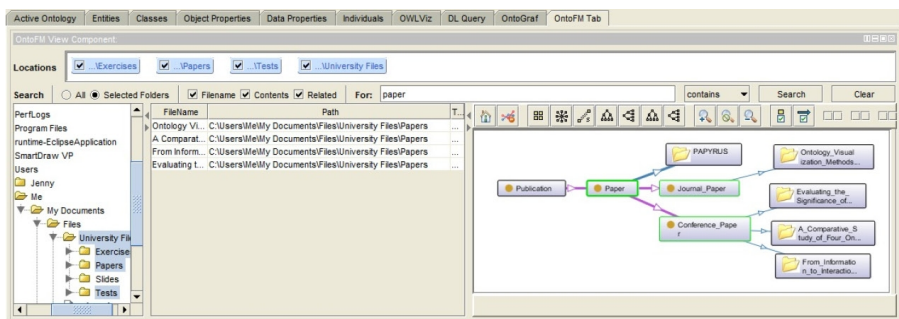

**Figure 2:** Searching for papers by limiting the semantic search to selected folders only. The entered term appears in the ontology visualization pane along with directly connected concept instances. The user may change the concept graph using the graph visualization controls on the visualization pane, and may expand/collapse concepts.

When the user searches for ontology keywords, the search is performed for all files in the computer. The user may select one or more folders from the folder hierarchy pane on the left to limit the search in those folders only. When this happens the "All" search selection automatically changes to "Selected Folders" updating the locations list on the top of the user interface with the names of the selected folders, as shown in Figure 2. In the case of a multiple folder selection without inputting any search terms, the ontology visualization pane shows all concepts related to the selected folders.

Time-related queries may be performed in two ways. The first is indirectly through the file list pane, when *all details* view is selected, by sorting files using the one of the time related columns (created or modified). The second is directly by specifying a constraint on the Date concept modeled in the ontology.

A short video demonstration and more screenshots of the OntoFM file manager are available in the OntoFM project page at http://www.uop.gr/~trifon/OntoFM/.

## 3 Outlook

We are planning to extend the functionality of the OntoFM file manager towards the maintenance and population of the personal ontology in a number of ways: (i) by letting users relate instances of concepts to files, groups of files, or folders, and (ii) by automatic metadata-based creation of concept instances for files or folders. Other extensions include weighting of concepts based on recent user activity as described in [7], viewing of relation types for files, and extensive user studies.

## References

[1] Beel, J. and Gipp, B., Enhancing search applications by utilizing mind maps. In HT 2010.
[2] Bernardi, A., Decker, S., van Elst, L., Grimnes, G., Groza, T., Handschuh, S., Jazayeri, M., Mesnage, C., Möller, K., Reif, G., and Sintek, M., The social semantic desktop: A new paradigm towards deploying the Semantic Web on the desktop. In SWEKS 2008.
[3] Chirita, A., Gavriloaie, R., Ghita, S., Nejdl, W., and Paiu, R., Activity based metadata for semantic desktop search, In ESWC 2005.
[4] Gauch, S., Chaffee, J., and Pretschner, A., Ontology-based personalized search and browsing. In WI AS 2003.
[5] Katifori, A., Halatsis, C., Lepouras, G., Vassilakis, C., and Giannopoulou, E., Ontology Visualization Methods - A Survey. In ACM Computing Surveys 2007.
[6] Katifori, A., Torou, E., Vassilakis, C., Lepouras, G., and Halatsis, C., Selected Results of a Comparative Study of Four Ontology Visualization Methods for Information Retrieval tasks. In RCIS 2008.
[7] Katifori, A., Vassilakis, C., and Dix, A., Using Spreading Activation through Ontologies to Support Personal Information Management. In CSKGOI 2008.
[8] Lepouras, G., Dix, A., Katifori, V., Catarci, T., Habegger, B., Poggi, A., and Ioannidis, Y., OntoPIM: From Personal Information Management to Task Information Management. In ACM SIGIR PIM Workshop 2006.
[9] Nepomuk wiki page accessible at http://nepomuk.semanticdesktop.org/
[10] Sauermann, L. and Heim, D., Evaluating long-term use of the Gnowsis Semantic Desktop for PIM. In ISWC 2008.